# Does the time horizon of the return predictive effect of investor sentiment vary with stock characteristics? A Granger causality analysis in the frequency domain

Yong Jiang[*], Zhongbao Zhou

School of Business Administration, Hunan University, Changsha 410082, China

**Abstract**

Behavioral theories posit that investor sentiment exhibits predictive power for stock returns, whereas there is little study have investigated the relationship between the time horizon of the predictive effect of investor sentiment and the firm characteristics. To this end, by using a Granger causality analysis in the frequency domain proposed by Lemmens et al. (2008), this paper examine whether the time horizon of the predictive effect of investor sentiment on the U.S. returns of stocks vary with different firm characteristics (e.g., firm size (Size), book-to-market equity (B/M) rate, operating profitability (OP) and investment (Inv)). The empirical results indicate that investor sentiment has a long-term (more than 12 months) or short-term (less than 12 months) predictive effect on stock returns with different firm characteristics. Specifically, the investor sentiment has strong predictability in the stock returns for smaller Size stocks, lower B/M stocks and lower OP stocks, both in the short term and long term, but only has a short-term predictability for higher quantile ones. The investor sentiment merely has predictability for the returns of smaller Inv stocks in the short term, but has a strong short-term and long-term predictability for larger Inv stocks. These results have important implications for the investors for the planning of the short and the long run stock investment strategy.

---

[*] Corresponding author, Email: jiangziya.ok@163.com..

# 1. Introduction

Behavioral theories posit that investor sentiment exhibits predictive power for stock returns, which has been extensively investigated in recent years (Baker and Wurgler, 2006, 2007; Stambaugh et al., 2012; Chung et al., 2012; Huang et al.,2015; Aloui et al., 2016; You et al., 2017). For example, Baker and Wurgler (2007) document that investor sentiment has a predictive power with respect to equity returns. Schmeling (2009) shows that the investors' sentiment acts as a significant predictor for stock returns for 18 industrialized countries. Stambaugh et al. (2012) find that investor sentiment is a significant negative predictor for the short legs of long-short investment strategies. Dergiades (2012) argues that the investor sentiment has a non-linear causality to stock returns by employing a nonlinear Granger causality model. Li et al. (2017) use a quantile Granger non-causality test model find a nonlinear causal relationship between investor sentiment and U.S. stock returns.

The common deficiency of the above-mentioned studies is that the causality assumption is limited to one particular data frequency, whereas they cannot analyze these frequency components separately, that is, the long-term components and the short-term components (see Breitung and Candelon, 2006; Lemmens et al., 2008). Consequently, they cannot identify whether the time horizon of predictive effect of investor sentiment on the returns of stock is in the short term or in the long term. Up to now, no literature examine whether the time horizon of the stock return predictive effect of investor sentiment varies for different firms.

In this paper, we address this issue by decomposing the Granger causality (GC) in the frequency domain, using frequency domain causality approach developed by Lemmens et al. (2008) based on spectral approach. The key idea of this approach is that a stationary process can be described as a weighted sum of sinusoidal components with a certain frequency. Instead of computing a single GC measure for the entire relationship, the GC is calculated for each frequency component separately. This analysis makes it possible to determine whether the predictive effect of investor sentiment is concentrated on short-term or long-term. To the best of our knowledge, the analysis of GC from investor sentiment to the returns of stocks with different levels of firm characteristics has not yet been explored in the frequency domain. By doing this, we provide evidence that the investor sentiment has different long-term and short-term predictive effects on stock returns by considering different levels of firm characteristics. (e.g., firm size, book-to-market equity rate, operating profitability and investment).

The paper proceeds as follows: Section 2 presents the data and methodology, Section 3 provides the empirical findings and Section 4 concludes.

**2. Data and methodology**

**2.1 Data**

The data are at a monthly frequency, spanning the period between July 1965 and September 2015. The US monthly investor sentiment index is taken from Baker and Wurgler (2007) [1]. The equally weighted portfolio returns formed on firm characteristics: firm size (Size), book-to-market equity rate (B/M), operating profitability (OP) and investment (Inv), which are collected from the website of Kenneth R. French[2].

**2.2 Frequency domain Granger causality test**

In this paper, we follow the bivariate GC test over the spectrum of Lemmens et al. (2008). Let $X_t$ and $Y_t$ be two stationary time series of length T. The goal is to test whether $X_t$ Granger-causes $Y_t$ at a given frequency $\lambda$. Pierce's (1979) measure for GC in the frequency domain is performed on the univariate innovations series, $\mu_t$ and $\nu_t$, derived from filtering the $X_t$ and $Y_t$ as univariate ARMA processes, which are white-noise processes with zero means, are possibly correlated with each other at different leads and lags.

Let $S_\mu(\lambda)$ and $S_\nu(\lambda)$ be the spectral density functions, or spectra, of $\mu_t$ and $\nu_t$ at a frequency $\lambda \in [0, \pi]$, defined by

$$S_\mu(\lambda) = \frac{1}{2\pi} \sum_{k=-\infty}^{\infty} \gamma_\mu(k) e^{-i\lambda k} \qquad (1)$$

$$S_\nu(\lambda) = \frac{1}{2\pi} \sum_{k=-\infty}^{\infty} \gamma_\nu(k) e^{-i\lambda k}, \qquad (2)$$

where $\gamma_\mu(k) = Cov(\mu_t, \mu_{t-k})$ and $\gamma_\nu(k) = Cov(\nu_t, \nu_{t-k})$ represent the autocovariances of $\mu_t$ and $\nu_t$ at a lag $k$. The idea of the spectral representation is that each time series may be decomposed into a sum of uncorrelated components, each related to a particular frequency $\lambda$. The spectrum can be interpreted as a decomposition of the series variance by frequency. The portion of the variance of the series occurring between any two frequencies is given by the area under the spectrum between those two frequencies. In other words, the area under $S_\mu(\lambda)$ and $S_\nu(\lambda)$ between any two frequencies $\lambda$ and $\lambda + d\lambda$, gives the portion of the variance of $\mu_t$ and $\nu_t$ respectively, due to cyclical components in the frequency band $(\lambda, \lambda + d\lambda)$.

---

[1] http://pages.stern.nyu.edu/~jwurgler/main.htm.
[2] http://mba.tuck.dartmouth.edu/pages/faculty/ken.french/index.html.

The cross spectrum represents the cross covariogram of two series in the frequency domain. It allows determining the relationship between two-time series as a function of frequency. Let $S_{\mu\nu}(\lambda)$ be the cross-spectrum between $\mu_t$ and $\nu_t$ series, which defined as

$$S_{\mu\nu}(\lambda) = C_{\mu\nu}(\lambda) + iQ_{\mu\nu}(\lambda) = \frac{1}{2\pi} \sum_{k=-\infty}^{\infty} \gamma_{\mu\nu}(k) e^{i\lambda k} \qquad (3)$$

where $C_{\mu\nu}(\lambda)$, called cospectrum and $Q_{\mu\nu}(\lambda)$, called quadrature spectrum are respectively, the real and imaginary parts of the cross-spectrum and $i = \sqrt{-1}$. Here $\gamma_{\mu\nu}(k) = Cov(\mu_t, \nu_{t-k})$ represents the cross-covariance of $\mu_t$ and $\nu_t$ at a lag $k$. The spectrum $Q_{\mu\nu}(\lambda)$ between the two series $\mu_t$ and $\nu_t$ at a frequency $\lambda$ can be interpreted as the covariance between the two series $\mu_t$ and $\nu_t$ that is attributable to cycles with frequency $\lambda$. The cross-spectrum can be estimated non-parametrically by

$$\hat{S}_{\mu\nu}(\lambda) = \frac{1}{2\pi} \left\{ \sum_{-M}^{M} w_k \hat{\gamma}_{\mu\nu}(k) e^{-i\lambda k} \right\} \qquad (4)$$

with the empirical cross-covariances $\hat{\gamma}_{\mu\nu}(k) = \hat{Cov}(\mu_t, \nu_{t-k})$ and window weights $w_k$, for $k = -M, \ldots, M$. Eq. (4) is called the *weighted covariance estimator*, and the weights $w_k$ are selected as the Bartlett weighting scheme i.e. $1 - |k|/M$. The constant $M$ determines the maximum lag order considered. The spectra of Eqs. (1) and (2) are estimated in a similar way. This cross-spectrum allows us to compute the coefficient of coherence $h_{\mu\nu}(\lambda)$ defined as

$$h_{\mu\nu}(\lambda) = \frac{|S_{\mu\nu}(\lambda)|}{\sqrt{S_\mu(\lambda) S_\nu(\lambda)}} \qquad (5)$$

Coherence can be interpreted as the absolute value of a frequency specific correlation coefficient, which takes values between 0 and 1. Lemmens et al. (2008)

have shown that under the null hypothesis that $h_{\mu\nu}(\lambda) = 0$, the estimated squared coefficient of coherence at the frequency $\lambda$., with $0 < \lambda < \pi$ when appropriately rescaled, converges to a chi-squared distribution with 2 degrees of freedom, denoted by $\chi_2^2$.

$$2(n-1)\hat{h}_{\mu\nu}^2(\lambda) \xrightarrow{d} \chi_2^2 \qquad (6)$$

where $\xrightarrow{d}$ stands for convergence in distribution, $n = T/(\sum_{k=-M}^{M} w_k^2)$. The null hypothesis $h_{\mu\nu}(\lambda) = 0$ versus $h_{\mu\nu}(\lambda) > 0$ is then rejected if

$$\hat{h}_{\mu\nu}(\lambda) > \sqrt{\frac{\chi_{2,1-\alpha}^2}{2(n-1)}} \qquad (7)$$

with $\chi_{2,1-\alpha}^2$ being the $1-\alpha$ quantile of the chi-squared distribution with 2 degrees of freedom. The coefficient of coherence in Eq. (5) gives a measure of the strength of the linear association between the two-time series, frequency by frequency, but does not provide any information on the direction of the relationship between the two processes. Lemmens et al. (2008) have decomposed the cross-spectrum (Eq. (1)) into three parts: (i) $S_{\mu \leftrightarrow \nu}$, the instantaneous relationship between $\mu_t$ and $\nu_t$; (ii) $S_{\mu \to \nu}$, the directional relationship between $\nu_t$ and lagged values of $\mu_t$; and (iii) $S_{\nu \to \mu}$, the directional relationship between $\mu_t$ and lagged values of $\nu_t$, i.e.,

$$S_{\mu\nu}(\lambda) = [S_{\mu \leftrightarrow \nu} + S_{\mu \to \nu} + S_{\nu \to \mu}] = \frac{1}{2\pi}\left[\gamma_{\mu\nu}(0) + \sum_{k=-\infty}^{-1} \gamma_{\mu\nu}(k)e^{-i\lambda k} + \sum_{k=1}^{\infty} \gamma_{\mu\nu}(k)e^{-i\lambda k}\right] \qquad (8)$$

The proposed spectral measure of GC is based on the key property that $\mu_t$ does not Granger-cause $\nu_t$ if and only if $\gamma_{\mu\nu}(k) = 0$ for all $k < 0$. The goal is to test the predictive content of $\mu_t$ relative $\nu_t$ to which is given by the second part of Eq. (8), i.e.,

$$S_{\mu \to \nu}(\lambda) = \frac{1}{2\pi}\left[\sum_{k=-\infty}^{-1} \gamma_{\mu\nu}(k)e^{-i\lambda k}\right] \qquad (9)$$

The Granger coefficient of coherence is then given by

$$h_{\mu \to \nu}(\lambda) = \frac{|S_{\mu \to \nu}(\lambda)|}{\sqrt{S_\mu(\lambda)S_\nu(\lambda)}} \qquad (10)$$

Therefore, in the absence of GC, $h_{\mu \to \nu}(\lambda) = 0$ for every $\lambda$ in $[0, \pi]$. The Granger coefficient of coherence takes values between zero and one (Pierce, 1979). Granger coefficient of coherence at frequency $\lambda$ is estimated by

$$\hat{h}_{\mu \to v}(\lambda) = \frac{|\hat{S}_{\mu \to v}(\lambda)|}{\sqrt{\hat{S}_{\mu}(\lambda)\hat{S}_{v}(\lambda)}} \quad (11)$$

with $|\hat{S}_{\mu \to v}(\lambda)|$ as in Eq. (4), but with all weights $w_k = 0$, for $k \geq 0$. The distribution of the estimator of the Granger coefficient of coherence is derived from the distribution of the coefficient of coherence (Eq. (6)). Under the null hypothesis $\hat{h}_{\mu \to v}(\lambda) = 0$, the distribution of the squared estimated Granger coefficient of coherence at frequency $\lambda$, with $0 < \lambda < \pi$ is given by

$$2(n'-1)\hat{h}^2_{\mu \to v}(\lambda) \xrightarrow{d} \chi^2_2 \quad (12)$$

where $n$ is now replaced by $n' = T/(\sum_{k=-M}^{-1} w_k^2)$. Since the weights $w_k$, with a positive index k, are set equal to zero when computing $\hat{S}_{\mu \to v}(\lambda)$, in effect only the $w_k$ with negative indices are taken into account. The null hypothesis $\hat{h}_{\mu \to v}(\lambda) = 0$ versus $\hat{h}_{\mu \to v}(\lambda) > 0$ is then rejected if

$$\hat{h}_{\mu \to v}(\lambda) > \sqrt{\frac{\chi^2_{2,1-\alpha}}{2(n'-1)}} \quad (13)$$

Afterward, we compute Granger coefficient of coherence given by Eq. (11) and test the significance of causality by making use of Eq. (13).

### 3. Empirical results

This section reports the results of causality tests in the frequency domain for two bivariate systems: investor sentiment and each stock returns of the portfolio for US. The variables have been filtered using ARMA models to obtain the innovation series. Both Augmented Dickey-Fuller (ADF) test (Dickey and Fuller, 1981) and Philip's Peron (PP) test (Phillips and Perron 1988) reject the null hypothesis of a unit root in all-time series at the 5% significance level.

**Table 1** Unit root and stationary test

|  | ADF | | PP | |
| --- | --- | --- | --- | --- |
|  | No trend | Trend | No trend | Trend |
| Sentiment | -3.447(4)*** | -3.412(4)*** | -3.427(12)** | -3.367(12)** |
| Size-small | -18.902(0)*** | -18.905(0)*** | -18.819(3)*** | -18.820(3)*** |
| Size-middle | -21.028(0)*** | -21.012(0)*** | -20.888(4)*** | -20.872(4)*** |
| Size-big | -22.281(0)*** | -22.263(0)*** | -22.237(5)*** | -22.219(5)*** |
| B/M-low | -22.733(0)*** | -22.716(0)*** | -22.726(5)*** | -22.698(4)*** |
| B/M-middle | -22.886(0)*** | -22.867(0)*** | -22.874(9)*** | -22.855(9)*** |
| B/M-high | -21.854(0)*** | -21.862(0)*** | -21.806(3)*** | -21.813(3)*** |
| OP-low | -22.093(0)*** | -22.078(0)*** | -22.084(4)*** | -22.070(4)*** |
| OP-middle | -23.062(0)*** | -23.043(0)*** | -23.053(7)*** | -23.034(7)*** |
| OP-high | -22.999(0)*** | -22.980(0)*** | -22.996(6)*** | -22.977(6)*** |
| Inv-small | -22.865(0)*** | -22.847(0)*** | -22.843(6)*** | -22.825(6)*** |
| Inv-middle | -23.355(0)*** | -23.337(0)*** | -23.378(8)*** | -23.360(8)*** |
| Inv-big | -22.371(0)*** | -22.352(0)*** | -22.354(3)*** | -22.335(3)*** |

Notes: **and ***indicate significance at the 5% and 1% level, respectively. The numbers in parentheses are the optimal lag order in the ADF and PP test based on the Schwarz Info criterion and Newey-west bandwidth.

In Fig.1, the estimated Granger coefficients of coherence are plotted versus the frequency $\lambda \in (0,\pi)$. This coefficient assesses whether, and to what extent, the investor sentiment is Granger causing the stock returns of the portfolio at that frequency. The higher the estimated Granger coefficient of coherence, the higher the Granger causality at that particular frequency. If the coefficient is higher than the 5% critical value, the investor sentiment is said to significantly "Granger cause" the stock returns at the frequency $\lambda$. Note that the lag length $M = \sqrt{T}$, the frequency can be translated into a cycle or periodicity of T months by $T = 2\pi/\lambda$, where T is the period. In this paper, we distinguish between the long-term components (low frequencies) and the short-term components (high frequencies) of a time series. We define the long-term components to have a cycle larger or equal to 12 months, which corresponds to the frequency $\lambda \leq 0.52$. The short-term components have a cycle smaller than 12 months, which corresponds to a frequency $\lambda > 0.52$.

As shown in Fig.1, for small Size stocks (bottom 30% quantiles), the null hypothesis of no causality is rejected for all frequencies at the 5% significance level, it indicates that the stock investor sentiment has highly significant predictive power for the stock returns in the short term and long term. However, for the big Size stocks (top 30% quantiles), the Granger coefficients of coherence corresponding to the low frequencies hardly reach statistical significance. It suggests that the stock investor sentiment have no predictability to the stock returns in long-term. Meanwhile, the null hypothesis should be rejected when $\lambda \in (2.5, 3)$ (cycles of 2.1–2.5 months), which reveals that investor sentiment Granger-causes stock returns of big Size stocks in the short term. This finding is in line with Lemmon and Portniaguina (2006) who find that investor sentiment can significant predict the returns of small size stocks.

Fig.1 shows that the investor sentiment is a poor predictor for the returns of smaller B/M stocks in long term but it has highly significant predictability in short term. And for bigger B/M stocks, the investor sentiment has not only short-term but also long-term predictability for returns. More specifically, we find that for smaller B/M stocks, when the frequency $\lambda > 0.52$, there exists a larger Granger coefficients of coherence in the range $\lambda \in (2.5, 3)$ which can reject the null hypothesis of no causality. It reveals that in the short term, the investor sentiment is a rich predictor of returns of smaller B/M stocks. However, when the frequency $\lambda \leq 0.52$, the null hypothesis cannot be rejected at the 5% significance level, which indicates that the investor sentiment doesn't have long-term predictability for the stock returns. With regard to bigger B/M stocks, it is found that the null hypothesis always can be rejected at the 5% significance level in low and high frequency. It confirms that the investor sentiment not only has short-term but also long-term predictability for the stock returns.

Fig.1 displays that the investor sentiment has predictability for the returns of lower OP stocks both in short term and in long term. However, the investor sentiment just has a predictability for the returns of larger OP stocks only in short term. In particular, for lower OP stocks, the Granger coefficients of coherence mostly are larger than the 5% critical value between the frequencies of 0 and $\pi$, which means that the investor sentiment has persistent predictability for the returns of lower OP stocks. However, for the higher OP stocks, it can reject the null hypothesis of no causality only when $\lambda$ is in the range $\lambda \in (2.4, 2.5)$ (cycles of 2.1–2.6 months), it reveals that investor sentiment only Granger-causes stock returns of higher OP stocks in the short term.

Next, we find that it has a different predictability from investor sentiment to the returns of stocks with low and high investment (see bottom panel of Fig.1). Specifically, it indicates that the investor sentiment is a poor predictor of returns of smaller Inv stocks in the long term but it has significant predictability in the short term. And for bigger Inv stocks, the investor sentiment not only has a short-term but also a long-term predictability for the stock returns.

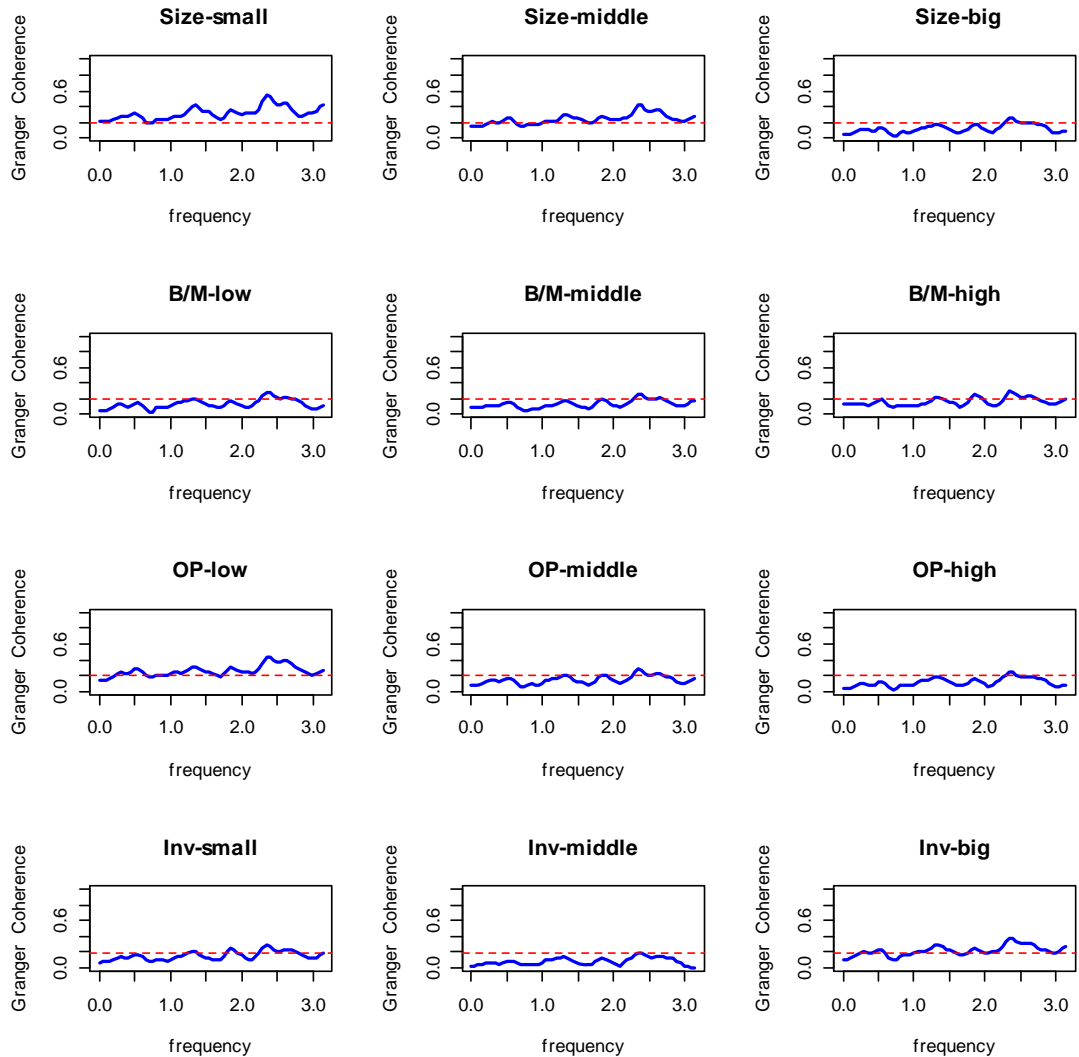

**Fig.1**. Granger coefficients of coherence for stock returns formed on Size, B/M, OP, Inv at 3 Deciles. The dashed line represents the critical value, at the 5% level, for the test for no Granger causality.

The paper considers the robustness of the results with respect to the stock returns of the portfolio at 10 Deciles (see Figs.2-5). It proves the robustness of our main results as follows.

Firstly, as shown in Fig.2, for the stock returns of portfolio formed on Size from Low10 to 8-Dec, it is proven that the Granger coefficients of coherence mostly are larger than the 5% critical value in all range $\lambda \in (0, \pi)$. However, for the stock returns of portfolio formed on Size in 9-Dec and Hi-10 deciles (bigger Size stocks), the null hypothesis of no causality can be rejected only when $\lambda$ is in the range $\lambda \in (2.5, 3)$. It is in line with our foregoing finding that the investor sentiment has strong predictability for the stock returns of smaller Size stocks both in the short term and long term, but for big stocks, only has a short term predictability.

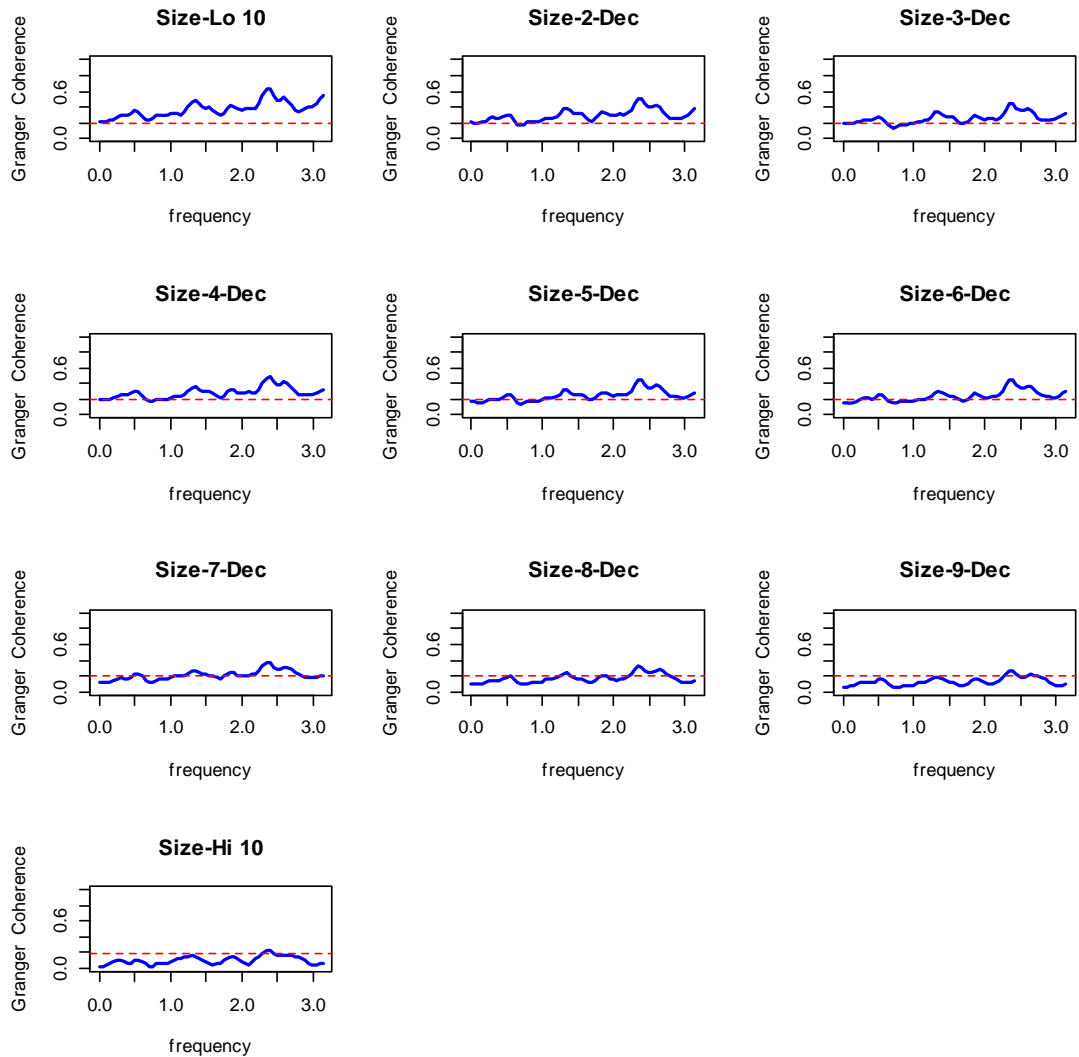

**Fig.2.** Granger coefficients of coherence for stock returns formed on Size at 10 Deciles. The dashed line represents the critical value, at the 5% level, for the test for no Granger causality.

Secondly, as shown in Fig.3, for the stock returns of portfolio formed on B/M from Low10 to 6-Dec deciles, it is found that the Granger coefficients of coherence value larger than the 5% critical value only when frequency $\lambda > 0.52$. For the higher B/M stocks (7-Dec, 8-Dec, 9-Dec and Hi-10 deciles), we find that it always has a larger Granger coefficient of coherence to reject the null hypothesis of no causality in the whole range $\lambda \in (0, \pi)$. It proved that the investor sentiment has significant predictability for the returns of lower B/M stocks both in the short term and long term, whereas only has a predictability for the returns of larger B/M stocks in the short term.

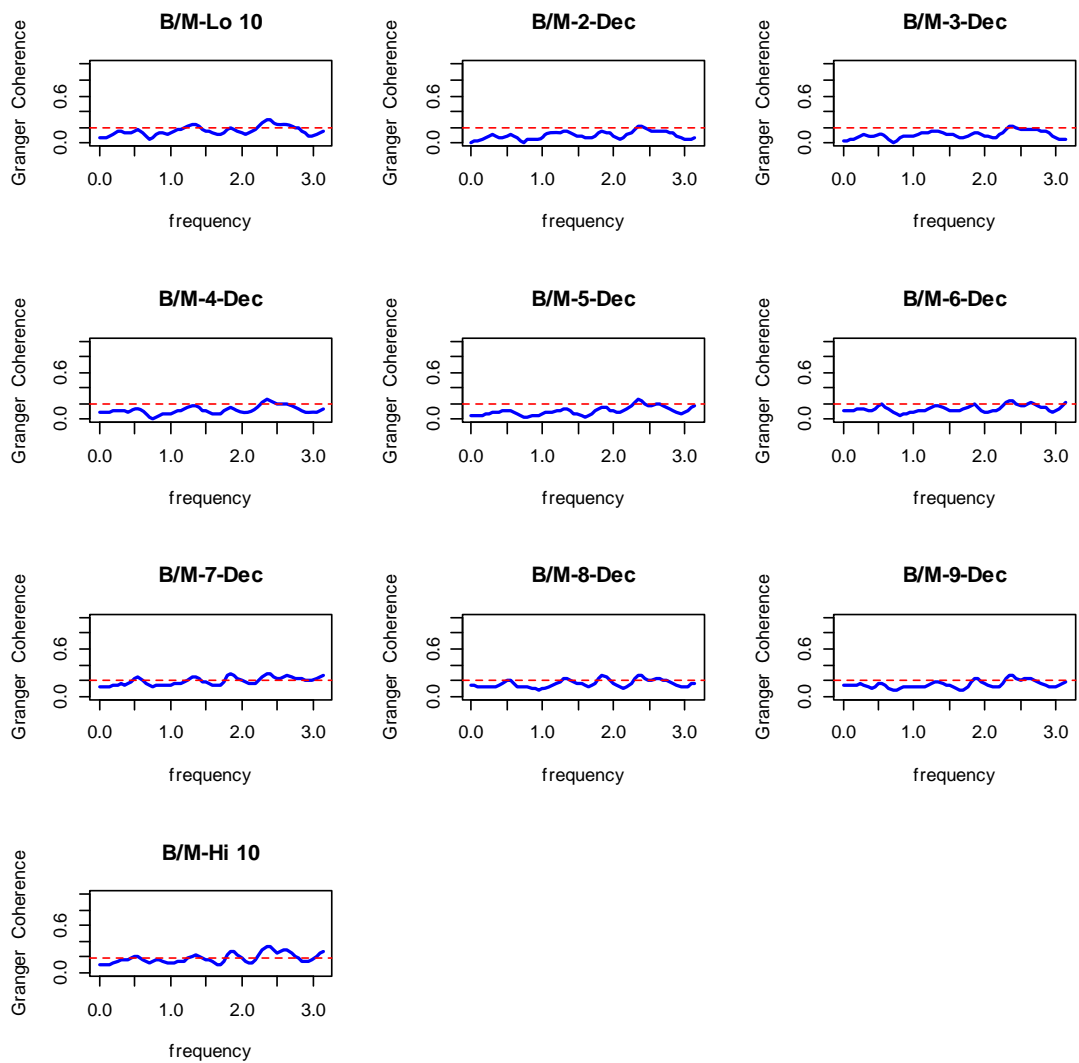

**Fig.3.** Granger coefficients of coherence for stock returns formed on B/M at 10 Deciles. The dashed line represents the critical value, at the 5% level, for the test for no Granger causality.

Thirdly, as shown in Fig.4, for the stock returns of portfolio formed on OP at Low10 and 2-Dec deciles (smaller OP stocks), it indicates that the null hypothesis of no causality mostly can be rejected when the frequency $\lambda > 0.52$ or $\lambda \leq 0.52$. For the higher OP stocks (3-Dec,4-Dec,5-Dec,6-Dec,7-Dec,8-Dec, 9-Dec, and Hi-10), it can reject the null hypothesis of no causality just when the frequency $\lambda > 0.52$. These findings assure that the investor sentiment has significant predictability for the returns of lower OP stocks both in the short term and long term, whereas only has a predictability for the returns of larger OP stocks in the short term.

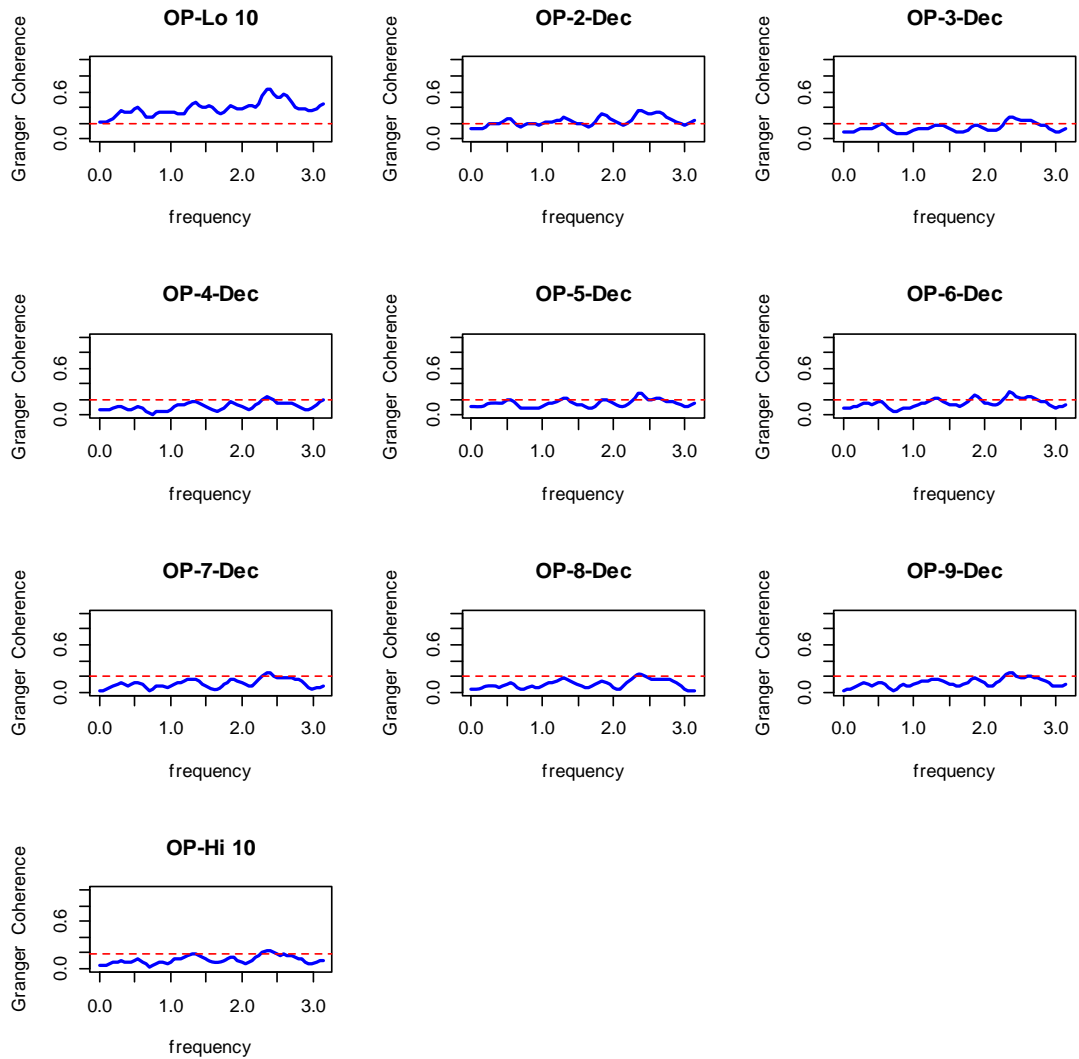

**Fig.4.** Granger coefficients of coherence for stock returns formed on OP at 10 Deciles. The dashed line represents the critical value, at the 5% level, for the test for no Granger causality.

Finally, as shown in Fig.5, for smaller Inv stocks (e.g. from 3-Dec to 8-Dec), the null hypothesis of no causality can be rejected only when the $\lambda$ is in the range $\lambda \in (2, 3)$ (cycles of 2.1-3.1 months). For bigger Inv stocks (9-Dec and Hi-10), it always can find higher Granger coefficients of coherence than the 5% critical value when the frequency $\lambda > 0.52$ or $\lambda \leq 0.52$. It confirms again that the investor sentiment only has predictability in the stock returns of smaller Inv stocks in the short term, but has a strong short term and long term predictability for larger Inv stocks.

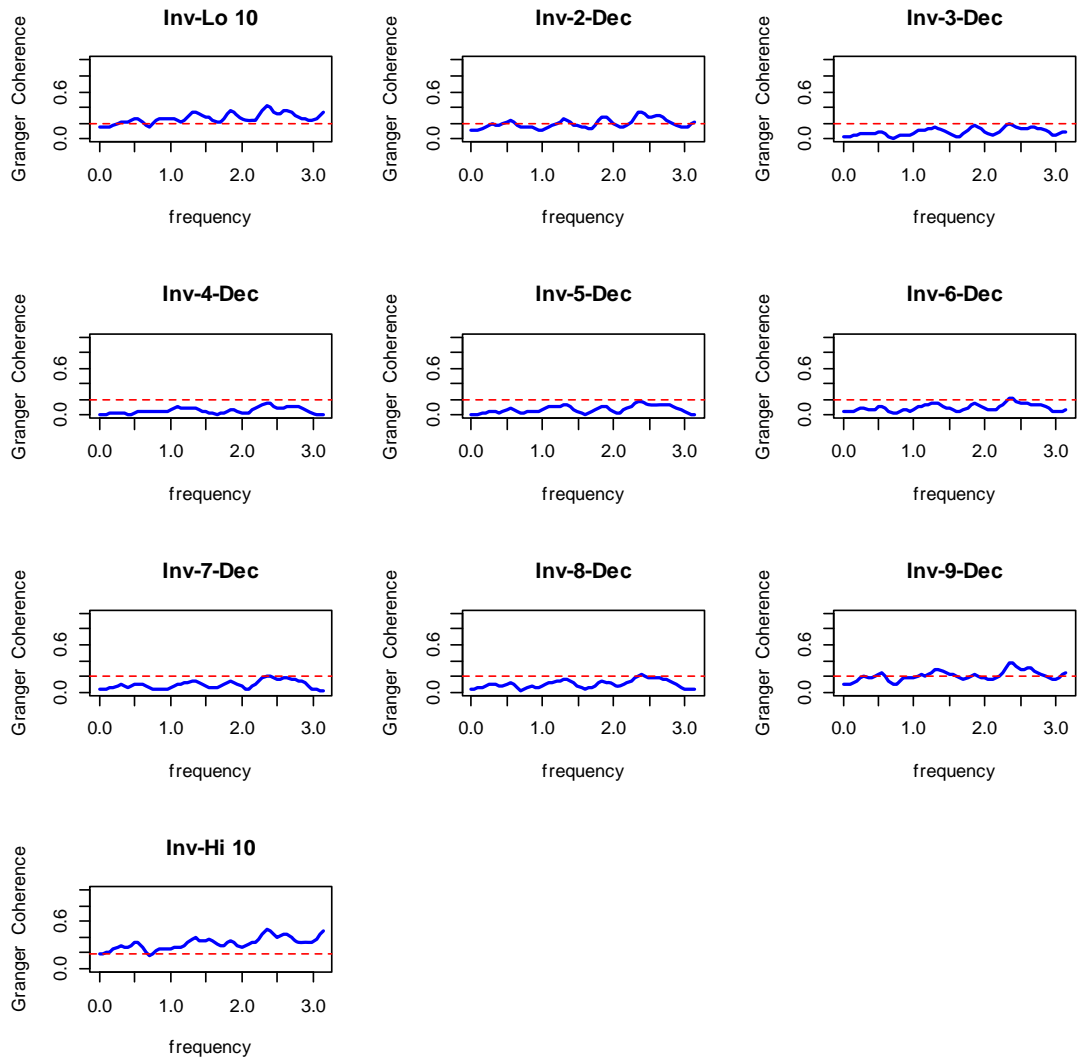

**Fig.5.** Granger coefficients of coherence for stock returns formed on Inv at 10 Deciles. The dashed line represents the critical value, at the 5% level, for the test for no Granger causality.

## 4. Concluding remarks

This paper aims to investigate whether the time horizon of return predictive effect of investor sentiment varies with stock characteristics (e.g., firm size, book-to-market equity rate, operating profitability and investment) by employing a Granger causality test in the frequency domain proposed by Lemmens et al. (2008). The evidence shows that investor sentiment has different predictive effects whose time horizon can be long-term or short-term on stock returns with different firm characteristics. More specifically, we find that 1) the investor sentiment has strong predictability in the stock returns for smaller Size stocks both in the short term and long term, but only has a short-term predictability for bigger Size stocks. 2) the investor sentiment has a significant predictive effect on the returns for lower B/M stocks both in the short term and long term, whereas only has a predictability for the returns of larger B/M stocks in the short term. 3) the investor sentiment has significant predictability for the returns of lower OP stocks both in the short term and long term, whereas only has a predictability for the returns of larger OP stocks in the short term. 4) the investor sentiment merely has predictability for the returns of smaller Inv stocks in the short term, but has a strong short-term and long-term predictability for larger Inv stocks. These results have important implications for the investors for the planning of the short and the long run stock investment strategy.

## Acknowledgments

We gratefully acknowledge the financial support from the National Natural Science Foundation of China (Nos. 71771082, 71431008) and Hunan Provincial Natural Science Foundation of China (No. 2017JJ1012).